\documentclass[aps,prl,twocolumn,10pt,longbibliography,nobibnotes]{revtex4-2}
\usepackage[utf8]{inputenc}
\usepackage[T1]{fontenc}

\usepackage{amsmath,amssymb,amsfonts,amsthm}
\usepackage{mathtools}
\usepackage{pifont}
\usepackage{bm}

\usepackage{graphicx}
\usepackage{dcolumn}
\usepackage{bm}
\usepackage{tikz} 
\usepackage{lipsum}
\usepackage{float}
\usepackage{subfigure}  
\usepackage{makecell}
\usepackage{tabularx} 

\usepackage{multirow}
\usepackage{booktabs}
\usepackage{amssymb}

\graphicspath{ {./Figs/} }
\definecolor{myurlcolor}{rgb}{0,0,0.7}
\definecolor{myrefcolor}{rgb}{0.8,0,0}
\usepackage{hyperref}
\hypersetup{
    unicode=true, bookmarksnumbered=false, bookmarksopen=false, breaklinks=false, pdfborder={0 0 0},
    colorlinks=true,
    linkcolor=myrefcolor,citecolor=myurlcolor,urlcolor=myurlcolor
}

\begin{document}

\newcommand{\bra}[1]    {\langle #1|}
\newcommand{\ket}[1]    {|#1 \rangle}
\newcommand{\ketbra}[2]{|#1\rangle\!\langle#2|}
\newcommand{\braket}[2]{\langle#1|#2\rangle}
\newcommand{\tr}[1]    {{\rm Tr}[ #1 ]}
\newcommand{\trr}[2]    {{\rm Tr}[ #1 ]_{\overline{#2}}}
\newcommand{\titr}[1]    {\widetilde{{\rm Tr}}\left[ #1 \right]}
\newcommand{\av}[1]    {\langle #1 \rangle}
\newcommand{\modsq}[1]    {| #1 |^2}
\newcommand{\0}    {\ket{\vec 0}}
\newcommand{\1}    {\ket{\vec 1}}
\newcommand{\dt}    {\delta\theta}
\newcommand{\I}    {\mathcal  I_{an}^{(q)}}
\newcommand{\Id}[1]    {\mathcal  I_q\left[#1\right]}
\newcommand{\Ic}[1]    {\mathcal  I_{\hat A}\left[#1\right]}
\newcommand{\C}    {\hat{\mathcal C}(t)}
\newcommand{\Cd}    {\hat{\mathcal C}^\dagger(t)}
\newcommand{\re}[1]    {\texttt{Re}\left[#1\right]}
\newcommand{\im}[1]    {\texttt{Im}\left[#1\right]}

\newcommand{\Enq}{\mathcal E_{N}^{(q)}}
\newcommand{\En}{\mathcal E_N}
\newcommand{\Q}{\mathcal Q}
\newcommand{\Qe}{\mathcal Q_{\rm ent}}
\newcommand{\Qb}{\mathcal Q_{\rm Bell}}
\newcommand{\Qep}[1]{\mathcal Q_{\rm ent}^{(p=#1)}}
\newcommand{\Qet}{\tilde{\mathcal Q}_{\rm ent}}
\newcommand{\Qbt}{\tilde{\mathcal Q}_{\rm Bell}}

\title{Generation and read-out of many-body Bell correlations with a probe qubit}

\author{Marcin P{\l}odzie\'n$^{1,2}$ and Jan Chwede\'nczuk$^{3}$}
\affiliation{$^1$
  Qilimanjaro Quantum Tech, Carrer de Veneçuela 74, 08019
Barcelona, Spain\\
$^2$ICFO-Institut de Ciencies Fotoniques, The Barcelona Institute of Science and Technology, 08860 Castelldefels (Barcelona), Spain\\
  $^3$ Faculty of Physics, University of Warsaw, ulica Pasteura 5, 02-093 Warszawa, Poland
  }

\begin{abstract}
  As demand for quantum technologies increases, so does the need to generate and classify non-classical correlations in complex many-body systems.
  We introduce a simple and versatile method for creating and certifying entanglement and many-body Bell correlations.
  This method relies on a single qubit interacting with an $N$-qubit system. We demonstrate that: (i) such pairwise interaction
  is sufficient to induce many-body quantum correlations, and (ii) the qubit can serve as a probe to extract all information about these correlations. 
  Thus, single-qubit measurements reveal multi-partite entanglement and $N$-body Bell correlations, enabling the rapid and efficient certification of non-classicality in complex systems.
  \end{abstract}

\maketitle

{\it Introduction.---}Significant technological advances followed the development of quantum theory. 
The invention of semiconducting transistors paved the way for the creation of personal computers and smartphones. 
Nowadays, these devices are connected by laser pulses propagating through waveguides, creating the Internet. 
Magnetic spin resonance scanners took the medical diagnoses to the next level. All of this and more was developed within the framework of \textit{the First Quantum Revolution}
which primarily utilizes single-body or collective non-classical effects~\cite{Aspect2024,Lambert2025}.
The \textit{Second Quantum Revolution}~\cite{Jaeger2018} builds on engineered many-body systems whose distinctive resources---quantum coherence~\cite{Streltsov2017},
quantum correlations/entanglement~\cite{ Horodecki2009, Horodecki2024} and Bell-type nonlocality~\cite{Brunner2014}---are absent in classical systems~\cite{DeChiara2018,Frerot2023,Laurell2024, Montenegro2025ManyBodyMetrology,Fazio2025OpenQMB}. Coherence supports controlled dynamics and interference. Meanwhile, quantum correlations offer advantages in nonclassical information processing, 
facilitating tasks like unconditionally secure communication~\cite{BennettBrassard1984,Ekert1991,LoChau1999,Gisin2002,Scarani2009,Xu2020}, 
and sub–shot noise metrology ~\cite{Giovannetti2006QuantumMetrology,Pezze2018RMP,Degen2017QuantumSensing,Agarwal2025,Montenegro2025}. 
For these purposes, quantum correlations must be reliably generated and certified.

The former task can be accomplished using the 
one-axis twisting (OAT) interaction of $N$ qubits. This method is useful for generating many-body Bell-correlated~\cite{Tura2014,Schmied2016,Aloy2019,Baccari2019,Tura2019,MullerRigat2021,Plodzien2022}
and spin-squeezed states, which find application in 
quantum metrology~\cite{Kitagawa1993,Wineland1994,Sorensen2001,esteve2008squeezing,Gross2010Nonlinear,Riedel2010AtomChip,Leroux2010CavitySqueezing,Norcia2018CavityOAT}.  

The latter requires more complex techniques, such as 
resource certification~\cite{PicenoMartnez2023,GuhneToth2009_EntanglementDetection,Peres1996,Horodecki1996,Toth2005_SpinModelWitnessPRA,Wiesniak2005_MagSusceptibilityNJP,Wiesniak2008}. 
Complementary methods include steering tests~\cite{Uola2020_SteeringRMP},
device-independent Bell tests and self-testing~\cite{Brunner2014_BellNonlocalityRMP,SupicBowles2020_SelfTestingReview}, 
measurement-device-independent witnesses~\cite{Branciard2013_MDIEW_PRL,Xu2014_MDIEW_Experiment}, 
as well as scalable reconstructions via compressed sensing and classical shadows~\cite{Gross2010,Huang2020}. 
Additionally, there are resource-specific tools, such as coherence witnesses~\cite{Streltsov2017_CoherenceRMP,Napoli2016_RobustnessOfCoherence,Ma2021_CoherenceWitnessesPRA}. 
As an alternative, the fidelity of a quantum state to the desired target can be estimated in order to certify 
quantum resources~\cite{Huang2025,HuangKuengPreskill2020, Aaronson2018ShadowTomography, ZhangSunFang2021PRL}.
Experiments have shown that quench protocol-based methods are a scalable way to certify quantum correlations~\cite{Hauke2016,Costa2021}.
Quench processes have also made it possible to directly measure R\'enyi entropies~\cite{Islam2015,Kaufman2016} and reconstruct entanglement growth over time using randomized measurements~\cite{Brydges2019}.
These processes also reveal the light-cone spreading of correlations, exposing  how entanglement is generated and transported in many-body systems~\cite{Cheneau2012}.

\begin{figure}[t!]
  \centering 
  \includegraphics[width=0.8\linewidth]{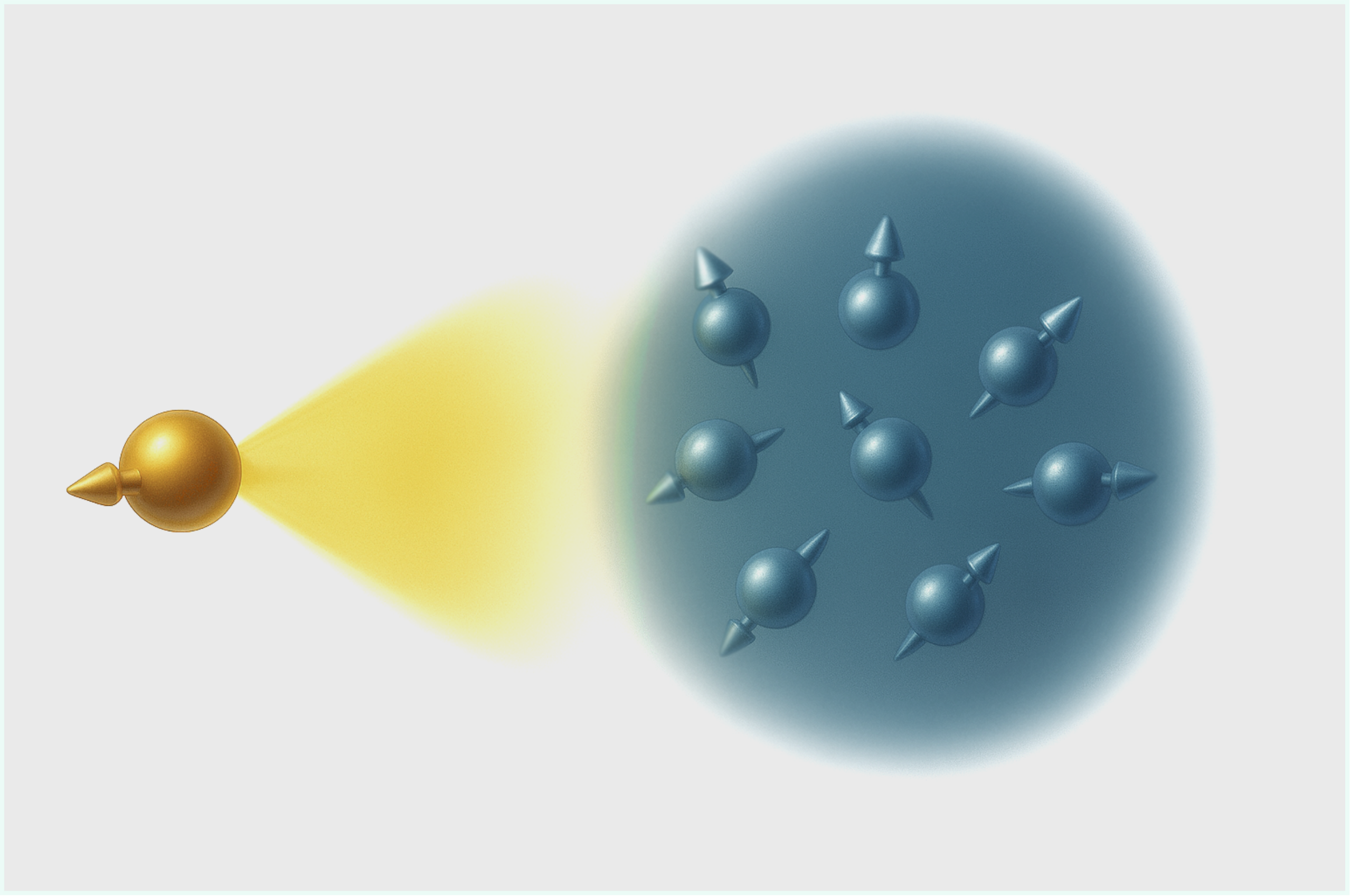}
  \caption{A visualization of a single-qubit probe ``observing'' a system of $N$ qubits.}
  \label{fig.scheme}
\end{figure}

In this work, we demonstrate that a single qubit interacting with an $N$-qubit ensemble can perform the aforementioned two tasks of generating and certifying quantum correlations. 
First, it can effectively initiate the OAT dynamics within the system. 
Second, we show that a single qubit can act as a probe to extract information from the system. Simple measurements on the probe enable us to verify the entanglement depth 
and many-body Bell correlations of a wide range of systems governed by Lipkin-Meshkov-Glick (LMG) Hamiltonians~\cite{Lipkin1965I, Lipkin1965II, Lipkin1965III}. 
These LMG models include OAT systems and describe interacting Bose-Einstein condensates (BECs) and long-range spin-1/2 chains in the lowest momentum mode. 
Additionally, they describe phonon- and photon-mediated interactions in spin-1/2 chains where the Dicke manifold is protected by an energy gap. 

A single qubit connected to a many-body system can reveal the characteristic function of the work distribution~\cite{PhysRevLett.110.230601}.
The probe qubit has been used to localize quantum critical points in Ising-type chains and to diagnose phase transitions by monitoring its coherence over time~\cite{Quan2006PRL,Zhang2008PRL}. 
It can detect nonergodicity and many-body localization in disordered spin chains~\cite{vanNieuwenburg2016PRB,Hetterich2018PRB} as well as determine ground-state phases and phase boundaries 
in the XXZ chain~\cite{Plodzien2025PRB}. Furthermore, it provides access to the generation of entanglement~\cite{Chiara_2006,Fogarty:12,PhysRevA.95.052132} and its spread across 
spatial cuts~\cite{SzaboTrivedi2022PRA,DoggenGornyiMirlin2024PRB,gessner2014local}. 
Meanwhile, qubit-only protocols determine whether a qubit becomes entangled with the environment during pure dephasing~\cite{Roszak2019PRA,Roszak2020PRR,Salamon2023,Strzalka2024PRA}. 
Little is known about using a single-qubit probe to generate and detect many-body quantum correlations. This work aims to bridge this gap by providing an experimentally friendly 
recipe for accomplishing these important steps for emerging quantum technologies.

We consider a state $\hat\varrho_N$ which depicts a collection of $N$ qubits---a natural platform
for many quantum-enhanced tasks. 
When an external probe qubit mediates interactions and generates many-body Bell correlations in the composite $N$-qubit system, 
the qubits are initially non-interacting and form a separable state. Next, we show that if these correlations already {\it exist}  in $\hat\varrho_N$, 
the probe can extract information about their strength and depth.

{\it Generation of correlations.---}We start with the {\it central-spin} Hamiltonian
{describing probe-qubit collectively interacting with the $(N-1)$-qubit system \cite{Shao2024} 
\begin{align}\label{eq.ham.int}
  \hat H =\frac{\Omega}{2}\,\hat\sigma^{(\mathrm{pr})}_z + \omega\,\hat J_z+g\big(\hat J_+\hat\sigma^{(\mathrm{pr})}_- + \hat J_-\hat\sigma^{(\mathrm{pr})}_+\big),
\end{align}
where $\hat\sigma^{(\mathrm{pr})}_z$ and $\hat\sigma^{(\mathrm{pr})}_{\pm}=\frac12(\hat\sigma^{(\mathrm{pr})}_x\pm i\hat\sigma^{(\mathrm{pr})}_y)$ are the probe-qubit operators.
The collective $(N-1)$-spin operator is 
$\hat J_z=\frac12\sum_{i=1}^{N-1} \hat\sigma_z^{(i)}$ and analogically for $\hat J_{\pm}$. 
In  the dispersive regime, when
$\Delta = \omega-\Omega$, $ |\Delta| \gg g$ and for as long as the evolution time $t$ satisfies $t\ll\chi^{-1}$, where $\chi=g^2/\Delta$, 
the second--order Schrieffer--Wolff (SW) transformation allows to express Eq.~\eqref{eq.ham.int} as~\cite{Shao2024,endm}
\begin{align}\label{eq.ham.oat}
  \hat H_{\rm OAT}\simeq\chi\hat S_z^2.
\end{align}
Here $\hat S_z=\hat J_z+\frac12\hat\sigma^{(\mathrm{pr})}_z$ is the composite spin operator. This is the OAT Hamiltonian that can correlate the $N$-qubit system. 
To characterize its potency and compare the exact evolution generated by the Hamiltonian~\eqref{eq.ham.int} with the effective OAT dynamics~\eqref{eq.ham.oat} we 
take a product state of $N$ qubits and the probe in the form $\ket{+1}_x^{\otimes N}$. Here, $\ket{+1}_x$ is a single-qubit eigenstate of $\hat\sigma_x$. Next, we 
evolve this state with either of the Hamiltonians and calculate the following correlator
\begin{align}\label{eq.corr.def}
  \Enq=\modsq{\av{\bigotimes_{k=1}^{N}\hat\sigma_{+}^{(k)}}},
\end{align}
Note that $\Enq\leqslant2^{-N}$ is an $N$-body Bell inequality, hence if 
$\mathcal Q_N=\log_2(\Enq2^{N})>0$, the system is Bell correlated, see Refs.
~\cite{ CavalcantiPRL2007,CavalcantiPRA2011,HePRA2011,Niezgoda2020,Niezgoda2021,Chwedeczuk2022,Plodzien2022,Plodzien2024PRA,Plodzien2024PRR,Hamza2024,Plodzien2025ROPP,Plodzien2025PRA,HernndezYanes2024,HernandezYanes2025} 
and~\cite{endm}.
In Fig.~\ref{fig.plots}(a) we plot $\mathcal Q_N$  as a function of the time normalized to the coupling strength, i.e., $\tau=2t\chi/\pi$. In this units, 
the dynamics is cyclic, with the period equal to unity. 
For both cases of Eq.~\eqref{eq.ham.int} and Eq.~\eqref{eq.ham.oat} we use  a total of $N=8$ particles. The
correlator in both cases is a very similar function of time. As expected from the validity of the SW transformation, the agreement is best for short times.
Most importantly, $\mathcal Q_N$ crosses the Bell threshold and reaches the maximal value $Q_N=N-2$ that is attainable only with the $N$-qubit Greenberger-Horne-Zeilinger (GHZ) state~\cite{endm}.
Therefore, we confirm that a single-qubit probe connected to an $N$-qubit ensemble via the Hamiltonian~\eqref{eq.ham.int} can closely mimic the OAT dynamics and generate many-body Bell correlations.

\begin{figure}[t!]
  \centering 
  \includegraphics[width=0.8\linewidth]{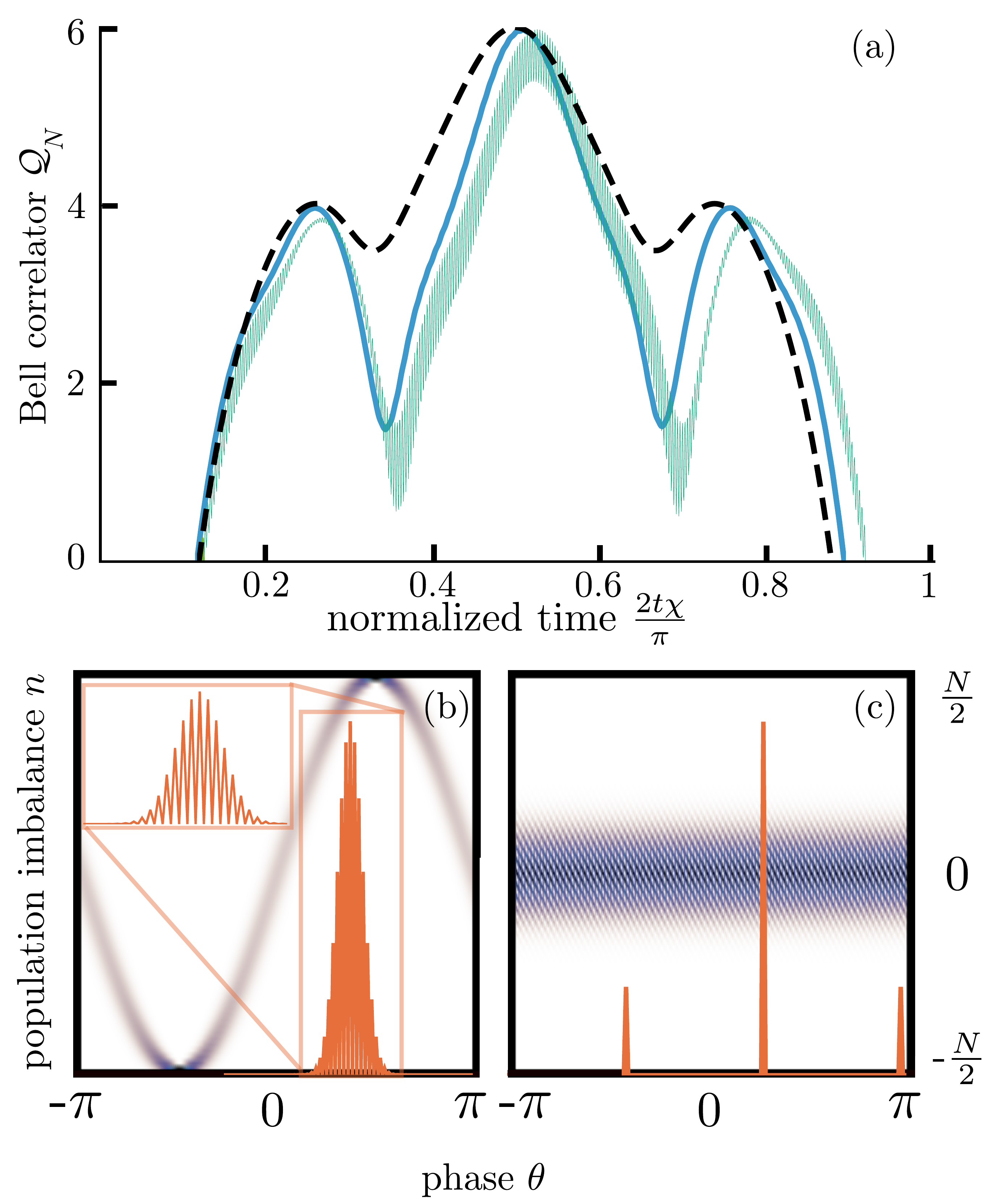}
  \caption{(a): Generation of many-body quantum correlations via single-qubit in central-spin model, Eq.~\eqref{eq.ham.int}. Lines present  the Bell correlator $\mathcal Q_N$ as a function of the 
    normalized time for $N=8$ qubits. The dashed black line denotes the ideal OAT scheme, while the solid blue is for $g=0.05$ 
    and the quickly oscillating green one uses $g=0.1$. For both cases, $\Omega=11$ and $\omega=1$. 
    (b) and (c): the main figures show the probabilities $p_n(\theta)$ as a function of $\theta$ and $n$ for the separable state (b) and the GHZ
    state (c), shown for $N=64$ using purple color. The presence of fine structures with respect to the variable $\theta$ in the latter case indicates the potential for sub-shot noise sensitivity.
      This is confirmed by the insets, which show the Fourier transform of $p_0(\theta)$. 
      The presence of the two side peaks in the latter case is indeed related to the fast variability of $p_0(\theta)$ and is a signature of Bell correlations.}
  \label{fig.plots}
\end{figure}

{\it Certification of correlations.---}We now demonstrate that a single-qubit probe can certify the presence of non-classical correlations embedded in $ \hat\varrho_N$.
To this end, the probe qubit governed by $\hat\varrho_{\mathrm{pr}}$
is ``connected'' to the system so that, initially, a product state is formed
\begin{align}\label{eq.couple}
  \hat\varrho=\hat\varrho_N\otimes\hat\varrho_{\mathrm{pr}}\,.
\end{align}
Next, the two parts interact in order to transfer the information about $\hat\varrho_N$ to the probe via  the Hamiltonian
\begin{align}\label{eq.ising}
  \hat H=\sum_{i=1}^NJ_i\hat\sigma_z^{(i)}\hat\sigma_z^{(\mathrm{pr})},
\end{align}
where $J_i$'s are the coupling strengths.
After time $t$ the density matrix of the probe becomes~\cite{endm} 
\begin{align}\label{eq.dens}
  \hat\varrho_{\mathrm{pr}}(\theta,t)=\left(\begin{array}{cc}
    \mathcal P & a \\
    a^* & 1-\mathcal P
  \end{array}\right).
\end{align}
The probability $\mathcal P$ does not depend on $t$, while
\begin{align}\label{eq.off}
  a=\sum_{\vec s=\pm1}\!p_{\vec s}\,e^{-2it\sum\limits_{i=1}^NJ_is_i}.
\end{align}
Here, the summation runs through $2^N$ elements of the eigen-basis of the $\hat\sigma_z^{(i)}$ operators and $\vec s=(s_1,\ldots,s_N)$ are their $\pm1$ eigen-values.
The diagonal element of the system's density matrix, $p_{\vec s}\equiv\varrho_N^{(\vec s,\vec s)}$ is the probability of the system occupying a state $\ket{\vec s}=\ket{s_1}\otimes\ldots\ket{s_N}$. 
If the system undergoes specific local operations prior to interacting with the probe, the element $a$ becomes sensitive to the off-diagonal elements of $\hat\varrho_N$ and 
provides information about its non-classicality.

For illustration, we pick a phase-$\theta$ imprint on each of $N$ qubits and the subsequent mixing of the signal using a $\pi/2$-pulse, namely
\begin{align}\label{eq.two.rots}
  \hat\varrho_N(\theta)=e^{-i\frac\pi2\hat J_x}e^{-i\theta\hat J_z}\hat\varrho_Ne^{i\theta\hat J_z}e^{i\frac\pi2\hat J_x}.
\end{align}
We now make a crucial observation: if the state
$\hat\varrho_N$ is permutationally invariant and all the couplings are equal, $J_i=J$, then the matrix element from Eq.~\eqref{eq.off} becomes
\begin{align}\label{eq.off.sym}
  a=\sum_{n=-\frac N2}^{\frac N2}\!p_n(\theta)\,e^{-i2\pi\frac{\tau}{N+1}n},
\end{align}
where $n=\frac12(n_{+1}-n_{-1})$ is the difference between the number of qubits occupying the $s=+1$ and the $s=-1$ state and $\tau=\frac{J}\pi(N+1)t$. 
Equation~\eqref{eq.off.sym} is a discrete Fourier transform of the probability $p_n(\theta)$ and its inversion yields
$p_n(\theta)$---a many-body probability from a measurement on a single-qubit probe. We will now demonstrate how to characterize non-classical correlations using the off-diagonal element $a$.

The first quantity derived from $p_n(\theta)$ that gives access to information on the multi-partite entanglement is 
the spin-squeezing parameter~\cite{Kitagawa1993,Wineland1994,Giovannetti2006QuantumMetrology}
\begin{align}\label{eq.squeeze}
  \xi^2=N\frac{\Delta^2\hat J_z(\theta)}{\left|\frac{\partial}{\partial\theta}\av{\hat J_z(\theta)}\right|^2},  
\end{align}
where both the numerator and the denominator can be obtained by calculating the two lowest moments of the probability, $\av{\hat J_z^\alpha(\theta)}=\sum_np_n(\theta)n^\alpha$ with $\alpha=1,2$.
When $\xi<1$, the system is entangled and useful for quantum metrology~\cite{Giovannetti2006QuantumMetrology}. 
Furthermore, from the value of $\xi$ one can deduce the entanglement depth of the system~\cite{Toth2005_SpinModelWitnessPRA,esteve2008squeezing,Hyllus2012PRA}. 

For non-Gaussian states that cannot be characterized only by the mean and the variance of $p_n$, higher moments can be calculated as in~\cite{Lucke2011}. 
The complete probability yields the Fisher information~\cite{Smerzi2014}
\begin{align}\label{eq.fi}
  \mathcal I=\sum_{n=-\frac N2}^{\frac N2}\frac{1}{p_n(\theta)}\left(\frac{\partial p_n(\theta)}{\partial\theta}\right)^2.
\end{align}
Finally, according to~\cite{Wysocki2025}, the off-diagonal element $a$ provides a lower bound for $\Id{\varrho_N}$---the quantum Fisher information (QFI) 
associated with the system's state~\cite{Braunstein1994}, namely
\begin{align}\label{eq.full}
  \Id{\hat\varrho_N(\theta)}\geqslant4\left(\frac{\re{\dot ae^{-i\phi}}^2}{1-|a|^2}+\im{\dot ae^{-i\varphi}}^2\right),
\end{align}
where $\varphi=\arg(a)$. The QFI sets the ultimate precision 
$\Delta\theta$ of phase estimation that can be achieved with $\hat\varrho_N(\theta)$. According to~\cite{Wysocki2025}, the above bound is tight for a large family
of relevant multi-qubit states. 

The above set of quantities, which all can be obtained from $a$, form a hierarchy
\begin{align}
  \frac{N}{\xi^2}\leqslant\mathcal I\leqslant\Id{\hat\varrho_N(\theta)}\leqslant(\Delta\theta)^{-2}.
\end{align}
Depending on the setting, any of these parameters can be used to determine the lower bound of the system's entanglement depth and metrological usability.

Next, we will demonstrate how to use the probe to detect the system's many-body Bell correlations, which are quantified by Eq.~\eqref{eq.corr.def}. 
The Bell correlator is characterized by a single element of the density matrix~\cite{endm}, $\modsq{\varrho_{\frac N2,-\frac N2}}$, that governs the GHZ coherence of the term
 $\ketbra{-1}{+1}^{\otimes N}$.
To access this element using $p_n(\theta)$, we note that this probability, according to Eq.~\eqref{eq.two.rots}, is
\begin{align}\label{eq.prob.th}
  p_n(\theta)=\sum_{m,m'=-\frac N2}^{\frac N2}\!\!\!\!d^*_{nm'}d_{mn}\varrho_{mm'}e^{-i(m'-m)\theta},
\end{align}
where $d_{mn}$ is a matrix element of the mixing operator $e^{-i\frac\pi2\hat J_x}$. The Fourier transform $\mathcal F\left[p_n(\theta)\right]$ 
with respect to $\theta$ reveals terms oscillating at different frequencies. 
In particular, the fastest-varying contribution to $\mathcal F_{\rm max}\left[p_n(\theta)\right]$ comes from the element of the above sum when $m'=-m=\frac N2$.
It is multiplied by the corresponding
matrix element $\varrho_{\frac N2,-\frac N2}$. On the other hand, the Bell correlator from Eq.~\eqref{eq.corr.def} is given by the same element
of the density matrix, see the discussion above Eq.~\eqref{eq.element}. Using the symmetry of the rotation matrix $d_{n,\frac N2}=d_{-\frac N2,n}$ and the expression
for the Bell correlator from Eq.~\eqref{eq.element}, we obtain that
\begin{align}
  \Enq[\hat\varrho_N]=\modsq{\varrho_{\frac N2,-\frac N2}}=\mathcal N\frac{\modsq{\mathcal F_{\rm max}\left[p_n(\theta)\right]}}{|d_{n,\frac N2}|^4},
\end{align}
where $\mathcal N$ is the normalization of the discrete Fourier transform. For most numerical procedures it is equal to $n_\theta^{-1}$, where 
  $n_\theta$ is the number of grid points on which $\theta$ is sampled.
This way, the Bell correlations, but also the $k$-locality and $k$-separability~\cite{Plodzien2025PRA} 
can be deduced from the measurements of a single-qubit probe.
The main part of Fig.~\ref{fig.plots} (b) and (c) shows the probability $p_n(\theta)$, plotted in purple, 
  as a function of $\theta$ and $n$ for two different pure states $\hat\varrho_N=\ketbra{\psi_N}{\psi_N}$. One is a separable state
\begin{align}
  \ket{\psi_N}=\left(\frac{\ket\uparrow+\ket\downarrow}{\sqrt2}\right)^{\otimes N}
\end{align}
while the other is the GHZ state
\begin{align}
  \ket{\psi_N}=\frac{\ket\uparrow^{\otimes N}+\ket\downarrow^{\otimes N}}{\sqrt2}
\end{align}
Clearly, the $p_n(\theta)$ for the former is a slowly-varying function of $\theta$, while the latter, reveals fine structures, characteristic for strongly entangled states~\cite{pezze2009entanglement}.
These rapid oscillations, due to a large derivative of $p_n(\theta)$ with respect to $\theta$, result in high metrological usability due to large Fisher information, see Eq.~\eqref{eq.fi}.
The Fourier transforms of $p_0(\theta)$ with respect to $\theta$ are shown for both cases as insets. The emergence  of two side-band peaks is a direct and experimentally accessible
confirmation of the GHZ-like coherence that drives the Bell correlations~\cite{Niezgoda2021}. 

{\it Applications.}---This method is ideal for characterizing 
multi-qubit states described by LMG models~\cite{Lipkin1965I, Lipkin1965II, Lipkin1965III}.
Their ground, thermal, and dynamically generated states all support many-body quantum correlations~\cite{Vidal2004, Latorre2005, Orus2008, SenDe2011, Du2011, Wang2012, ZhangLiPLA2013, Fadel2018, Loureno2020, Bao2020,  BaoEtAlPRA2020, LourencoEtAl2020,GlobalQDLMGJPCM2021, Hengstenberg2023, Loureno2025,  Plodzien2024PRA}. 
LMG models have been implemented on different experimental platforms, including superconducting qubits 
and BECs~\cite{Loureno2025, Morrison2008, Chen2009,Larson2010,Grimsmo2013,Gelhausen2018,Hobday2023,Sauerwein2023}.
Furthermore, LMG systems are well-suited for studying quantum phase transitions and the geometric phase can be deduced from the probability
$p_n(\theta)$~\cite{CUI2006243,Chen:09,PhysRevE.94.032123,PhysRevB.103.174104}.

Moreover, the proposed protocol is useful for more than just initializing the OAT dynamics; it can also be used to detect entanglement and Bell correlations in such systems. 
The OAT can be generated using a two-component Bose-Hubbard model~\cite{Sorensen1999,Kajtoch2018}, as well as in spinful Fermi-Hubbard model with external lasers~\cite{HernandezYanes2022}. 
In both scenarios, the probe should interact with the system's zero mode, resulting in collective coupling.  
The proposed protocol can be also utilized in short-range spin chains, in particular the staggered XXX model that realizes the OAT dynamics~\cite{Gietka2020, HernandezYanes2022}.  
In Rydberg tweezer arrays, an XXX chain can be engineered by tuning microwave-programmable XXZ interactions~\cite{Scholl2022_PRXQ_XXZ}
or by realizing Heisenberg-type exchange via off-resonant dressing in 1D chains~\cite{Wang2024_PRX_AnisoH}.
Other applications are the OAT systems based on the resonant dipole–exchange evolution accompanied by the Floquet/control protocols 
to balance the couplings~\cite{DeLeseleuc2017_OpticalControlResDipPRL,Nishad2023_FloquetSpinExchangePRA,Zhao2023_FloquetTailoredNatComm,Carrera2025}
or on dual-species architectures and long-lived circular Rydberg implementations~\cite{Anand2024_DualSpeciesRydberg,Semeghini2025_CircularRydbergPRXQ}. 
In trapped-ion systems, global M\o{}lmer–S\o{}rensen and geometric-phase gates can be  used to realize collective, 
effectively all-to-all OAT dynamics~\cite{Sorensen1999_PRL_ThermalMotion, Sorensen2000_PRA_ThermalMotion, Sackett2000_Nature_4IonGHZ, SchmidtKaler2003_Nature_CZCNOT, 
  Leibfried2003_Nature_GeomPhaseGate, Roos2004_Science_3Qubit, Haeffner2005_Nature_WStates, Leibfried2005_Nature_SixCat}.

{\it Discussion.}---The single-qubit probe gives access to the distribution of $N$ parties, $p_n(\theta)$, among their two modes. Naturally, $p_n(\theta)$ could be accessed by directly performing
measurements on each of the qubits. While such on-site measurements are feasible in chain/lattice configurations~\cite{sherson2010single}, they become much harder
when the system is a Bose gas~\cite{Lucke2011,Smerzi2014}. Hence the method presented here, requiring a measurement on only one qubit, can significantly facilitate
the detection of many-body correlations in complex systems. 

In principle, it can also serve for remote tomography of $N$-body permutation-symmetric states. 
The computational complexity of the tomography of such states scales like $N^2$, since it uses collective measurements~\cite{Toth2010PI,Moroder2012PI,Schwemmer2014Comp}.
On the contrary, the measurement of the entanglement depth and the veryfication of metrological usability scales linearly with $N$ in our protocol, 
as required by the inversion of the discrete Fourier transform in Eq.~\eqref{eq.off.sym}. 
Another O($N$) number of points is necessary for detecting Bell correlations [see Eq.~\eqref{eq.prob.th}].
Therefore, the number of measurements is the same in both cases, and no additional detection noise is introduced.
Furthermore, direct tomography requires accurate collective  detection prone to measurement errors that scale with $N$.
By contrast, single-qubit tomography that yields access to the off-diagonal element $a$, is elementary~\cite{James2001Qubit,Hradil1997QSE,Rehacek2001Iterative,James2001Qubit}.

{\it Conclusions.}---In this work we have shown that many-body quantum correlations, such as entanglement or Bell correlations, can be generated and
certified using a single-qubit probe that interacts with the system. The effective OAT dynamics that stems from a coupling of the probe to each of the qubits yields a complex Bell-correlated system.
Furthermore, quantities like the spin-squeezing or the (quantum) Fisher information can be derived from a one-body measurement and provide information about the entanglement depth and
the presence of correlations useful for quantum metrology. Moreover, a many-body Bell correlator can also be deduced within this scheme, giving access to the most non-classical quantum correlations.
The single-qubit probe can also detect the quantum phase transition in the ground state of a broad family of LMG systems.
This protocol competes with other methods of generating and detecting quantum many-body effects by eliminating the need for cumbersome many-body measurements.

{\it Acknowledgements.---}This work was supported by the National Science Centre, Poland, within the QuantERA II Programme that has received funding from the European Union’s Horizon 2020 
research and innovation programme under Grant Agreement No 101017733, Project No. 2021/03/Y/ST2/00195.
M.P. acknowledges support from: European Research Council AdG NOQIA; MCIN/AEI (PGC2018-0910.13039/501100011033, CEX2019-000910-/10.13039/501100011033, Plan National FIDEUA PID2019-106901GB-I00, Plan National STAMEENA PID2022-139099NB, I00, project funded by MCIN/AEI/10.13039/501100011033 and by the “European Union NextGenerationEU/PRTR" (PRTR-C17.I1), FPI); QUANTERA DYNAMITE PCI2022-132919, QuantERA II Programme co-funded by European Union’s Horizon 2020 program under Grant Agreement No 101017733;
Ministry for Digital Transformation and of Civil Service of the Spanish Government through the QUANTUM ENIA project call - Quantum Spain project, and by the European Union through the Recovery, Transformation and Resilience Plan - NextGenerationEU within the framework of the Digital Spain 2026 Agenda;
Fundació Cellex; Fundació Mir-Puig; Generalitat de Catalunya (European Social Fund FEDER and CERCA program; Funded by the European Union. Views and opinions expressed are however those of the author(s) only and do not necessarily reflect those of the European Union, European Commission, European Climate, Infrastructure and Environment Executive Agency (CINEA), or any other granting authority. Neither the European Union nor any granting authority can be held responsible for them (HORIZON-CL4-2022-QUANTUM-02-SGA PASQuanS2.1, 101113690, EU Horizon 2020 FET-OPEN OPTOlogic, Grant No 899794, QU-ATTO, 101168628), EU Horizon Europe Program (This project has received funding from the European Union’s Horizon Europe research and innovation program under grant agreement No 101080086 NeQSTGrant Agreement 101080086 — NeQST); ICFO Internal “QuantumGaudi” project.

The data that support Fig.~\ref{fig.plots} are openly available~\cite{data}.

\section{END MATTER}

\setcounter{equation}{0}                                                                                                
\renewcommand{\theequation}{EM.\arabic{equation}}                                                                        

\subsection{Central-spin model to OAT}

In the following we outline the second--order Schrieffer--Wolff (SW) transformation
that maps the isotropic central--spin Hamiltonian to an effective one--axis twisting (OAT) form. 
We consider $\hat H = \hat H_0 + \hat V$, where 
\begin{align}
  \hat H_0 &=\frac{\Omega}{2}\,\hat\sigma^{(\mathrm{pr})}_z + \omega\,\hat J_z\\
  \hat V&=g\big(\hat J_+\hat\sigma^{(\mathrm{pr})}_- + \hat J_-\hat\sigma^{(\mathrm{pr})}_+\big),
\end{align}
which uses the notation from the main text.
We focus on the dispersive regime
$\Delta = \Omega - \omega$, $ |\Delta| \gg g$,
so that direct spin exchange is off--resonant.  
Projectors onto the probe-spin manifolds are
\begin{align}
  \hat P = \frac{1-\hat\sigma^{(\mathrm{pr})}_z}{2}, \qquad \hat Q = \frac{1+\hat\sigma^{(\mathrm{pr})}_z}{2}, \qquad \hat P + \hat Q = \mathbb I .
\end{align}
To get rid of the  off–diagonal couplings between low- and high-energy subspaces at first order,
we perform a Schrieffer–Wolff transformation
$\hat H_{\rm eff}=e^{\hat S}\hat H e^{-\hat S}$ with an anti-Hermitian generator $\hat S$.

At the leading order we choose $\hat S$ to eliminate the $\mathcal O(g)$ $\hat P$–$\hat Q$ couplings, i.e. we fix the first–order generator
$\hat S^{(1)}$ to satisfy $[\hat H_0,\hat S^{(1)}]=-\,\hat V$, which yields
\begin{equation}
\hat S^{(1)}=\frac{g}{\Delta}\big(\hat J_+\hat\sigma^{(\mathrm{pr})}_- - \hat J_-\hat\sigma^{(\mathrm{pr})}_+\big). 
\end{equation}
Expanding $e^{\hat S}\hat H e^{-\hat S}$ to second order gives
\begin{equation}
\hat H_{\mathrm{eff}}^{(2)}=\hat H_0+\frac{1}{2}\,[\hat S^{(1)},\hat V].
\end{equation}
With $\hat X=\hat J_+\hat\sigma^{(\mathrm{pr})}_-$ and $\hat V=g(\hat X+\hat X^\dagger)$, one finds
\begin{equation}
\frac{1}{2}\,[\hat S^{(1)},\hat V]=\frac{g^2}{\Delta}\,[\hat X,\hat X^\dagger]. 
\end{equation}
Using $\hat J_\pm \hat J_\mp=\hat J^2-\hat J_z^2\pm \hat J_z$,
$\hat\sigma^{(\mathrm{pr})}_-\hat\sigma^{(\mathrm{pr})}_+=\tfrac{1}{2}(1-\hat\sigma^{(\mathrm{pr})}_z)$, and
$\hat\sigma^{(\mathrm{pr})}_+\hat\sigma^{(\mathrm{pr})}_-=\tfrac{1}{2}(1+\hat\sigma^{(\mathrm{pr})}_z)$,
we obtain
\begin{equation}
  [\hat X,\hat X^\dagger]=\hat J_z-\hat\sigma^{(\mathrm{pr})}_z\big(\hat J^2-\hat J_z^2\big). 
\end{equation}
The effective Hamiltonian becomes
\begin{equation}
  \hat H_{\mathrm{eff}}^{(2)}=\frac{\Omega}{2}\,\hat\sigma^{(\mathrm{pr})}_z+\omega\,\hat J_z+\frac{g^2}{\Delta}\,\hat J_z-\frac{g^2}{\Delta}\,\hat\sigma^{(\mathrm{pr})}_z\big(\hat J^2-\hat J_z^2\big).
\end{equation}
Projecting onto the $\hat P$ sector, $\hat H_{\downarrow}
=\hat P \hat H_{\mathrm{eff}}^{(2)} \hat P$, yields
\begin{equation}
\hat H_{\downarrow}
=-\frac{\Omega}{2}
+\Big(\omega+\frac{g^2}{\Delta}\Big)\hat S_z
+\frac{g^2}{\Delta}\hat S^2
-\frac{g^2}{\Delta}\hat S_z^2,
\end{equation}
where $\hat S_i$ is the collective $N$-spin/probe operator.
Within the symmetric manifold ($S=(N+1)/2$), $\hat S^2$ is a constant and can be dropped.
Defining $\tilde\omega\equiv \omega+\chi$ and $\chi\equiv \frac{g^2}{\Delta}$,
we  
\begin{equation}
\hat H_{\downarrow}=\tilde\omega\,\hat S_z-\chi\,\hat S_z^2.
\end{equation}
Finally, by moving to the rotating frame via $\hat U_z(t)=e^{+i\tilde\omega t \hat S_z}$, we obtain an effective OAT model
\begin{equation}
\hat H_{\rm coll}=\hat U_z\hat H_{\downarrow}\hat U_z^\dagger-i\hat U_z\partial_t\hat U_z^\dagger
= -\,\chi\,\hat S_z^2 .
\end{equation}
Accuracy requires $g/|\Delta|\ll 1$ and evolution times $t\ll |\Delta|/g^2$ so that $\mathcal O(g^3/\Delta^2)$ terms remain negligible.

In the dispersive regime $\epsilon g/|\Delta|\ll1$, the SW mapping reproduces the short-time growth of $\mathcal Q_N$, as aniticipated [see. Fig.~\ref{fig.plots}(a)]. 
However,  it remains accurate even in the vicinity of the GHZ peak (at $\tau=1/2$) because the accumulated contribution of the leading neglected higher-order terms scales as 
  $g/(|\Delta|\tau)$ and for the considered parameters it is small at this time.

\subsection{Bell correlator}

A many-body Bell inequality
is constructed under the assumption of binary outcomes of measurements of two quantities $\sigma^{(k)}_{x/y}=\pm1$, where  $k\in\{1\ldots N\}$ labels 
the particles~\cite{ CavalcantiPRL2007,CavalcantiPRA2011,HePRA2011,Niezgoda2020,Niezgoda2021,Chwedeczuk2022,Plodzien2022,Plodzien2024PRA,Plodzien2024PRR,Hamza2024,Plodzien2025ROPP,Plodzien2025PRA,HernandezYanes2025}.
If the $N$-body correlation function $\En\equiv\modsq{\av{\prod_{k=1}^N\sigma_+^{(k)}}}$ is consitent with the postulates of local realism, it satisfies
\begin{align}\label{eq.lhv.2}
  \En&=\modsq{\int\!\!d\lambda\,p(\lambda)\prod_{k=1}^N\sigma_+^{(k)}(\lambda)}\nonumber\\
  &\leqslant\int\!\!d\lambda\,p(\lambda)\prod_{k=1}^N\modsq{\sigma_+^{(k)}(\lambda)}=2^{-N},
\end{align}
where $p(\lambda)$ is the probability density of a hidden variable, $\sigma_{+}^{(k)}=\frac12(\sigma_x^{(k)}+i\sigma_y^{(k)})$ and
in the last step we used the Cauchy-Schwarz inequality and the fact that $\modsq{\sigma_+^{(k)}(\lambda)}=1/2$.
This is the many-body Bell inequality as its violation defies the postulate of local realism.
For qubits $\sigma_+^{(k)}$'s are replaced with the rising Pauli operators and the Bell inequality becomes
\begin{align} 
  \Enq[\hat\varrho_N]=\modsq{{\rm Tr}\bigg[{\hat{\varrho}_N\,\bigotimes_{k=1}^N\hat\sigma_+^{(k)}}\bigg]}\leqslant2^{-N}.
\end{align}
The above product of $N$ rising operators is non-zero only if it acts on $\ket{-1}^{\otimes N}$, giving $\ket{+1}^{\otimes N}$. Therefore
$\Enq$ is a measure of coherence between these two, which is maximal for the GHZ state, $\ket{\psi}=\frac1{\sqrt2}(\ket{-1}^{\otimes N}+\ket{+1}^{\otimes N})$,
and can be expressed as follows
\begin{align}\label{eq.element}
  \Enq[\hat\varrho_N]=\modsq{\varrho_{\frac N2,-\frac N2}}.
\end{align}
Here, $\modsq{\varrho_{\frac N2,-\frac N2}}$ denotes the matrix element of $\hat\varrho_N$ multiplying $\ketbra{-1}{+1}^{\otimes N}$.

\subsection{Expression for the antenna's density matrix}\label{sec.gen}

The general expression for the $N$-qubit state is
\begin{align}
  \hat\varrho_N=\sum_{\vec s,\vec s'=\pm1}\varrho_{\vec s,\vec s'}(\theta)\ketbra{\vec s}{\vec s'}.
\end{align}
Its coupling to the single-qubit probe as in Eq.~\eqref{eq.couple} gives
\begin{align}
  \hat\varrho=\sum_{\vec s,\vec s'=\pm1}\sum_{s_p,s_p'=\pm1}\varrho_{\vec s,\vec s'}\varrho_{s_p,s_p'}\ketbra{\vec s}{\vec s'}\otimes\ketbra{s_p}{s_p'}.
\end{align}
The subsequent time-evolution generated by the Hamiltonian~\eqref{eq.ising} yields
\begin{align}
  \hat\varrho(t)&=\sum_{\vec s,\vec s'=\pm1}\sum_{s_p,s_p'=\pm1}\varrho_{\vec s,\vec s'}\varrho_{s_p,s_p'}\ketbra{\vec s}{\vec s'}\otimes\ketbra{s_p}{s_p'}\times\nonumber\\
  &\times e^{-it\sum\limits_{i=1}^NJ_i(s_is_p-s'_is'_p)}.
\end{align}
Once we trace-out the $N$-qubits' degrees of freedom we are left with
\begin{align}
  \hat\varrho_{\mathrm{pr}}(t)&=\sum_{s_p,s_p'=\pm1}\varrho_{\vec s,\vec s}\varrho_{s_p,s_p'}\ketbra{s_p}{s_p'}e^{-it\sum\limits_{i=1}^NJ_is_i(s_p-s'_p)}
\end{align}
The off-diagonal term is obtained when setting $s_p=1=-s_p'$ (or vice-versa), which yields Eq.~\eqref{eq.off}.

\end{document}